\documentclass[11pt]{article}
\usepackage{moriond,epsfig}
\usepackage{graphicx,mathrsfs}

\def\be{\begin{equation}}
\def\ee{\end{equation}}
\def\bea{\begin{eqnarray}}
\def\eea{\end{eqnarray}}

\begin{document}
\vspace*{4cm}
\title{THE SUSY FLAVOR PROBLEM IN 5D GUTS}

\author{ S. FICHET }

\address{Laboratoire de Physique Subatomique et Cosmologie,\\
53, rue des Martyrs, Grenoble, France}

\maketitle\abstracts{
In 5D SUSY GUTs, wave-function localization permits to reproduce flavour hierarchy. As this mechanism also acts on SUSY breaking
parameters, it can potentially solve the SUSY flavour problem. We carry out an analysis of the Holographic Grand Unification framework,
where we take properly into account effects of matrix anarchy. In this contribution, we focus on brane-localized SUSY breaking and its consequences.
%which is a warped $SU(6)$ GUT with pNGB Higgses on the IR brane. SUSY is also broken on the IR brane. 
%We modelize matrix anarchy
%in a proper way to make quantitative statements 
}

\section{Flavour hierarchies in 5 dimensions\label{sec:1}}
%\subsection{Flavour puzzles}\label{subsec:1}

In the Standard Model (SM), the three generations of quarks and leptons follow a peculiar pattern of hierarchical masses and mixings. On the other hand, models with TeV-scale supersymmetry (SUSY) generically induce large, unobserved flavour violating neutral currents (FCNCs) through their scalar SUSY breaking soft terms. It is temptating 
to assume that the mechanism solving the Standard Model flavour puzzle also gives a peculiar structure to scalar soft terms, such that large FCNCs are suppressed.

An attractive mechanism permitting to realize this idea is wave-function localization \cite{Grossman:1999ra}.
Indeed, localising the Standard Model matter fields in the bulk of a compact extra dimension, for instance on a slice of AdS$_5$ \cite{Randall:1999ee}, naturally leads to such flavour hierarchies.
Depending on localization of Higgs fields and supersymmetry breaking, this localization can also alleviate the SUSY flavour problem (see, for example, \cite{Choi:2003fk,Nomura:2008pt,Dudas:2010yh}). Supersymmetry breaking can take place on a brane, or in the gravitational background. We will
here consider the former possibility, for a more general analysis, see \cite{WIP}. In that case, by localizing the Higgs fields and SUSY breaking on the same brane, the soft terms  follow a similar hierarchy structure as the Yukawa couplings.%, and will thus be approximately diagonal in the fermion mass eigenstate basis. 
%A related mechanism was originally advocated in a 4D setup, with the visible sector fields acquiring large anomalous dimensions from their couplings to a strongly coupled CFT \cite{Nelson:2000sn}. These two pictures ca be argued to be related by AdS/CFT duality \cite{Choi:2003fk}. 

\subsection{A 5D realization}

\begin{figure}
%\rule{5cm}{0.2mm}\hfill\rule{5cm}{0.2mm}
%\vskip 2.5cm
%\rule{5cm}{0.2mm}\hfill\rule{5cm}{0.2mm}
\centering
\includegraphics[width=12cm,trim = 0cm 8cm 0cm 5cm, clip]{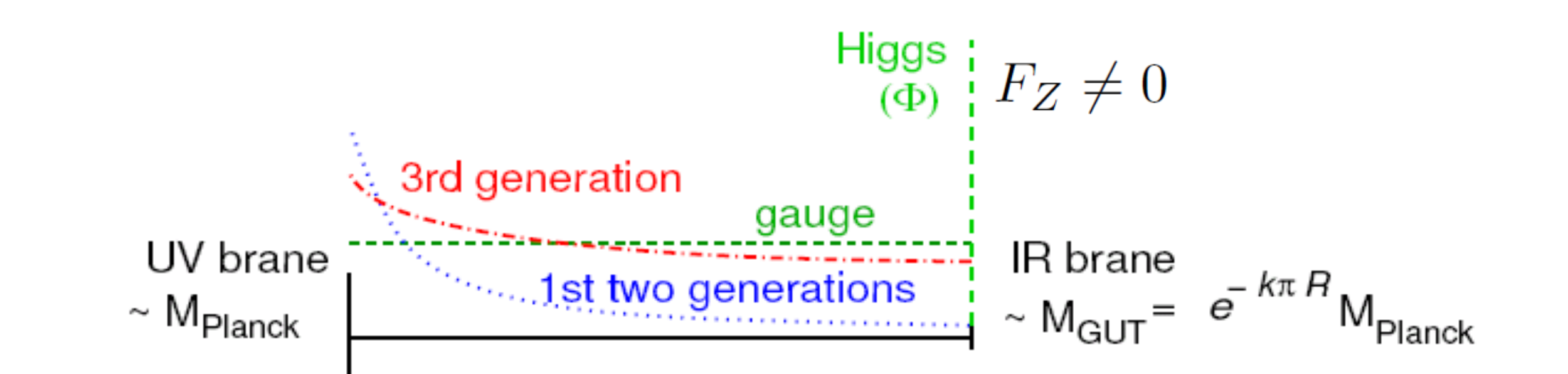}
\caption{The 5D framework considered. Higgs and the SUSY breaking fields are localized on the IR brane. Matter fields propagate in the bulk, with different exponential profiles generating flavour hierarchy.
\label{fig:sketch_HGU}}
\end{figure}

Most of our analysis is sufficiently broad to cover, at least quantitavely,  any 5D SUSY GUT with the Higgs and SUSY breaking sectors localised on the same brane.  However, when we do need to work with a concrete model for definiteness, we choose the ``holographic GUT'' model of Nomura, Poland and Tweedie (NPT) \cite{Nomura:2006pn}. A basic picture is given on Figure \ref{fig:sketch_HGU}. In the NPT model, there is a warped extra dimension, and the bulk gauge group is $SU(6)$. It is broken by boundary conditions to $SU(5)\times U(1)$ on the UV brane, and by the VEV of an adjoint brane field $\Sigma$ to $SU(4)\times SU(2)\times U(1)$ on the IR brane. This gives essentially the Standard Model gauge group in the 4D effective field theory. Matter fields are localised in the bulk, and boundary conditions are chosen such that their zero modes furnish precisely the matter content of the MSSM.% The MSSM Higgs fields are pseudo-Goldstone bosons arising from $\Sigma$ \cite{Inoue:1985cw,Contino:2003ve}, and the symmetry breaking structure results in a GUT-scale degenerate Higgs mass matrix \cite{Brummer:2010gh}, thus automatically solving the $\mu$ problem. Models of this type have been shown to be able to give realistic low-energy mass spectra when assuming flavour-blind supersymmetry breaking \cite{Brummer:2010gh}. 

Within this 5D framework, the 4D effective Yukawa couplings generated by the overlap of matter fields with Higgs fields are: 
\be
Y_{u}=\left(\begin{array}{ccc}
\varepsilon^{4} & \varepsilon^{3} & \varepsilon^{2}\\
\varepsilon^{3} & \varepsilon^{2} & \varepsilon\\
\varepsilon^{2} & \varepsilon & 1\end{array}\right)~,~~~~~
Y_{d}=Y_{l}^{t}=\varepsilon\left(\begin{array}{ccc}
\varepsilon^{2} & \varepsilon^{2} & \varepsilon^{2}\\
\varepsilon & \varepsilon & \varepsilon\\
1 & 1 & 1\end{array}\right)~.\ee%~~~~~~
%V_{CKM}\sim\left(\begin{array}{ccc}
%1 & \varepsilon & \varepsilon^{2}\\
%\varepsilon & 1 & \varepsilon\\
%\varepsilon^{2} & \varepsilon & 1\end{array}\right)\]
The scalar supersymmetry breaking parameters are 
\be
A_{u,d,l}\sim\frac{F_{Z}}{M_{*}}Y_{u,d,l}~,~~~
m_{Q,U,E}^{2}\sim\left|\frac{F_{Z}}{M_{*}}\right|^{2}\left(\begin{array}{ccc}
\varepsilon^{4} & \varepsilon^{3} & \varepsilon^{2}\\
\varepsilon^{3} & \varepsilon^{2} & \varepsilon\\
\varepsilon^{2} & \varepsilon & 1\end{array}\right)~,~~~
m_{D,L}^{2}\sim\left|\frac{F_{Z}}{M_{*}}\right|^{2}\varepsilon^{2}\left(\begin{array}{ccc}
1 & 1 & 1\\
1 & 1 & 1\\
1 & 1 & 1\end{array}\right)~.\ee
$M^*$ is the 5D cutoff scale, and $F^Z/M^*$ is the SUSY scale.
Note that this SUSY GUT is $SU(5)$-like,  Yukawa couplings and soft masses thus satisfy 
$SU(5)$ relations. These relations are however only approximate due to $SU(5)$ breaking operators residing on
the IR brane.

%After discussing the scalar soft terms, what about gaugino masses ? 
%We assume  universal gaugino masses $M_a=M_{1/2}$ . Provided that there is some theoretical uncertainty, and as these soft terms are crucial
%for phenomenology, we simply choose  a generic parametrization with a coefficient $\alpha_{1/2}$: 

To work out the phenomenology, we also need to specify the gaugino masses. We assume universality: $M_a=M_{1/2}$, and choose a generic parametrization:
\be
M_{1/2}=\alpha_{1/2} \frac{F^Z}{M^*}~.
\ee

\subsection{Quantifying matrix anarchy}

The above mechanism  permits to elegantly explain flavour hierarchy. 
More precisely, it permits to transform anarchical matrices, whose  elements are all of same order of magnitude,
into hierarchical matrices, though multiplication by powers of $\varepsilon$.
The same situation also appears in the Frogatt-Nielsen mechanism.

Even if anarchy of the original flavour matrices is overwhelmed by $\varepsilon$ factors, it is  necessary to parametrize and quantify it properly.
Indeed, on one hand, some amount of matrix anarchy is still necessary to reproduce precisely the SM masses and CKM matrix.
On the other hand, this anarchy can introduce uncertainty in the SUSY spectrum, and in flavour observables.

We call the elements of the original anarchical flavour matrices $\lambda^{u,d,e}_{ij}$, such that $Y^{u,d,e}_{ij}\propto\lambda^{u,d,e}_{ij} \varepsilon^{n_{ij}}$, where $n_{ij}$ corresponds to the appropriate power of $\varepsilon$. These $\lambda$'s are complex,  $\mathcal{O}(1)$ coefficients. Other $\lambda$'s also appear in the scalar soft terms. %, there is thus $\mathcal{O}(100)$ 
Since  we do not study CP violation, we take them to be real without loss of generality.
But there is still a freedom on their $\pm$ signs. 
%One could neglect this freedom,
%by considering all coefficients positive, but this is in fact a very unatural choice.
%For example, with  all coefficients positive, two eigenvalues of each yukawa matrices are exactly zero.
Just taking all $\lambda$'s positive would be a very unnatural choice, as in that case 
two eigenvalues of each Yukawa matrices are exactly zero.
We do not restrict ourselves to an arbitrary choice of sign combinations, but instead scan over all physical, inequivalent combinations.
%By taking $\mathscr{L}\neq1$, these two eigenvalues will no longer be zero, but will have a widespread $\mathscr{L}$-dependent distribution. The predictivity is thus lost in that case, making the flavour model meaningless.
%We will thus consider only sign combinations leading  to three non zero eigenvalues in yukawa matrices.
%After discussing  the phases of the $\lambda$'s, what about their magnitudes ?
Regarding the magnitude of the $\lambda$'s, as they are multiplicative coefficients, it is natural
to let them vary within a range $[\mathscr{L}^{-1},\mathscr{L}]$, $\mathscr{L}$ being $\mathcal{O}(1)$. The logarithm of this range is symmetric, and it is in fact more intuitive to consider $\log |\lambda|$. The most natural probability density function associated to $\log |\lambda|$ (i.e. the prior) should be also symmetric, and we choose the simplest possible: the uniform distribution.
We thus have
\be
p(|\Lambda|=|\lambda|)=U(-\log\mathscr{L},\log\mathscr{L})~,
\ee
$U(a,b)$ being the uniform distribution on the interval $[a,b]$.
$\mathscr{L}=1$ corresponds to setting  all the $|\lambda|$'s to one, i.e to suppress  matrix anarchy in magnitude.

The zero eigenvalues which can appear in  Yukawa matrices for certain sign combinations are no longer zero once $\mathscr{L}\neq1$. 
They have instead a widespread $\mathscr{L}$-dependent distribution. The predictivity being lost in that case, we consider 
only sign combinations leading  to three non zero eigenvalues.
%Indeed, even if one neglects the phases of the $\mathcal{O}(1)$ coefficients, 
%there is still a freedom on the $\pm$ sign. One could neglect this freedom,
%by considering all coefficients positive, but this is in fact an unatural choice.
%The fact is , with  all coefficients positive, two eigenvalues of each yukawa matrices are exactly zero, keeping  $\mathscr{L}=1$.
%By taking $\mathscr{L}\neq1$, these two eigenvalues will no longer be zero, but will have a widespread $\mathscr{L}$-dependent distribution. The predictivity is thus lost in that case, making the flavour model meaningless.
%We will thus consider only sign combinations leading  to three non zero eigenvalues in yukawa matrices.

\section{Phenomenological aspects}

In the framework described above, we are left with four parameters: the SUSY scale $F^Z/M^*$, the gaugino mass parameter $\alpha_{1/2}$, the ratio of the two Higgs vevs $\tan\beta=v_u/v_d$, and the magnitude of flavour matrix anarchy $\mathscr{L}$. We emphasize that $\mathscr{L}$ should be considered as a parameter of the model.

\subsection{The lightest supersymmetric particle}

A crucial aspect of the SUSY spectrum is the nature of the lightest supersymmetric particle (LSP). 
%Assuming that it is the neutralino and that it is stable, 
%this is a good candidate for WIMP dark matter.  On the contrary, if a charged slepton is lighter than the neutralino, one has to assume that it decays into 
%a lighter SUSY particle like the gravitino or the axino. 
%So what is the LSP in our framework ?
%What is the LSP in our framework ?
%If one neglects the flavour matrix anarchy, i.e. fix $\mathscr{L}=1$, the LSP is a charged slepton, mostly right selectron.
In our framework, for $\mathscr{L}=1$, the LSP is a charged slepton, mostly right selectron.
However, with $\mathscr{L}>1$, the probability $\textrm{P}(\tilde{\chi}^0_1\textrm{LSP})$ to have a neutralino LSP becomes non zero. This can be understood
by considering  one-loop RGEs. 
%Indeed, the right selectron RGE is 
%$
%16\pi^{2}\frac{d}{dt}m_{e}^{2}=-6g_{2}^{2}\left|M_{2}\right|^{2}-\frac{6}{5}g_{1}^{2}\left|M_{1}\right|^{2}-\frac{3}{5}g_{1}^{2}S~.$
%It depends on the RG invariant 
%$
%S=m_{H_u}^2-m_{H_d}^2+\textrm{Tr}(m^2_Q-m^2_L-2m^2_U+m^2_D+m^2_E)~.
%$
%For $\mathscr{L}=1$, $S$ is exactly zero due to $SU(5)$ relations between soft masses. But with $\mathscr{L}>1$, the cancellations are not exact anymore, 
%and this modifies the running of $m_{e}^{2}$.
Indeed, for $\mathscr{L}=1$, the RG invariant $S=m_{H_u}^2-m_{H_d}^2+\textrm{Tr}(m^2_Q-m^2_L-2m^2_U+m^2_D+m^2_E)$ is exactly zero due to $SU(5)$ relations between soft masses. But when $\mathscr{L}>1$, the cancellations are not exact anymore, and $S$ modifies the running of the selectron mass.

\begin{figure}
\centering
\includegraphics[width=4.8cm,trim = 0cm 3.5cm 10cm 15.3cm, clip]{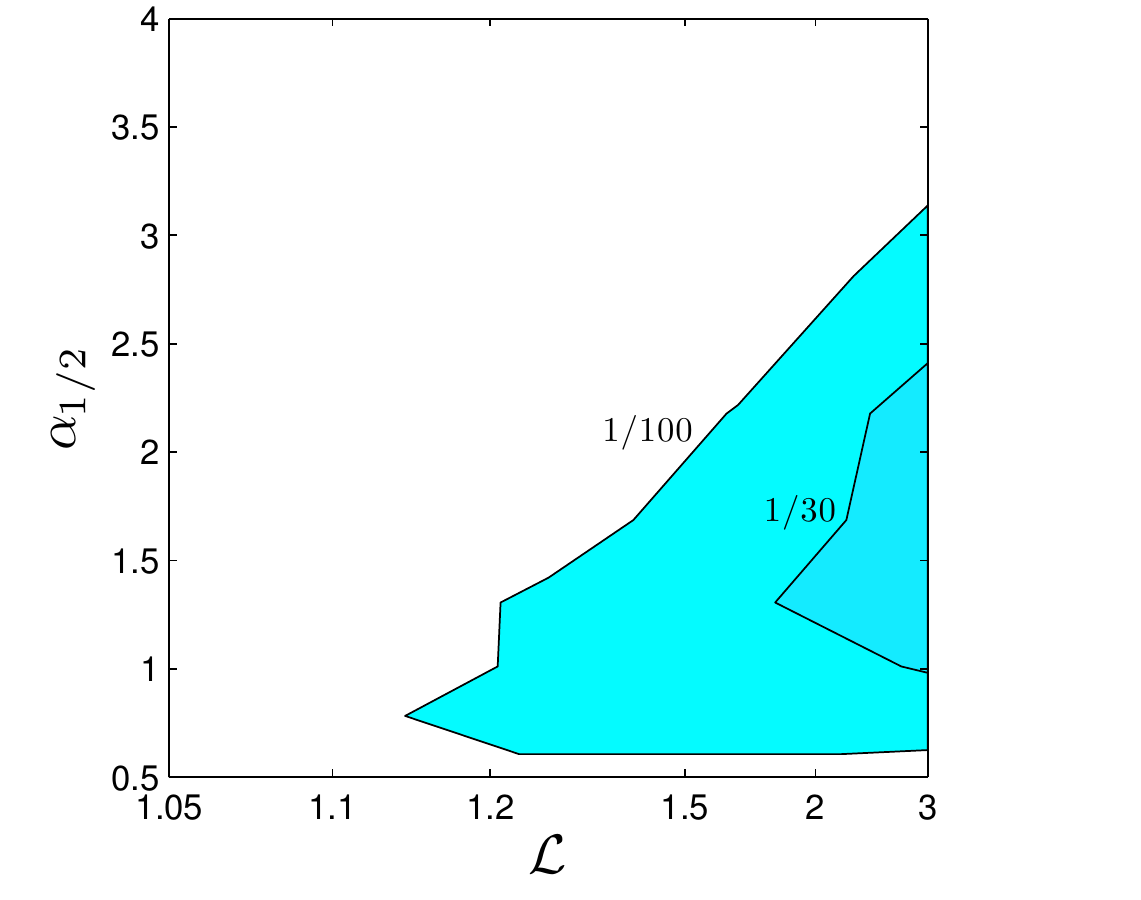}
\includegraphics[width=4.8cm,trim = 0cm 3.5cm 10cm 15.3cm, clip]{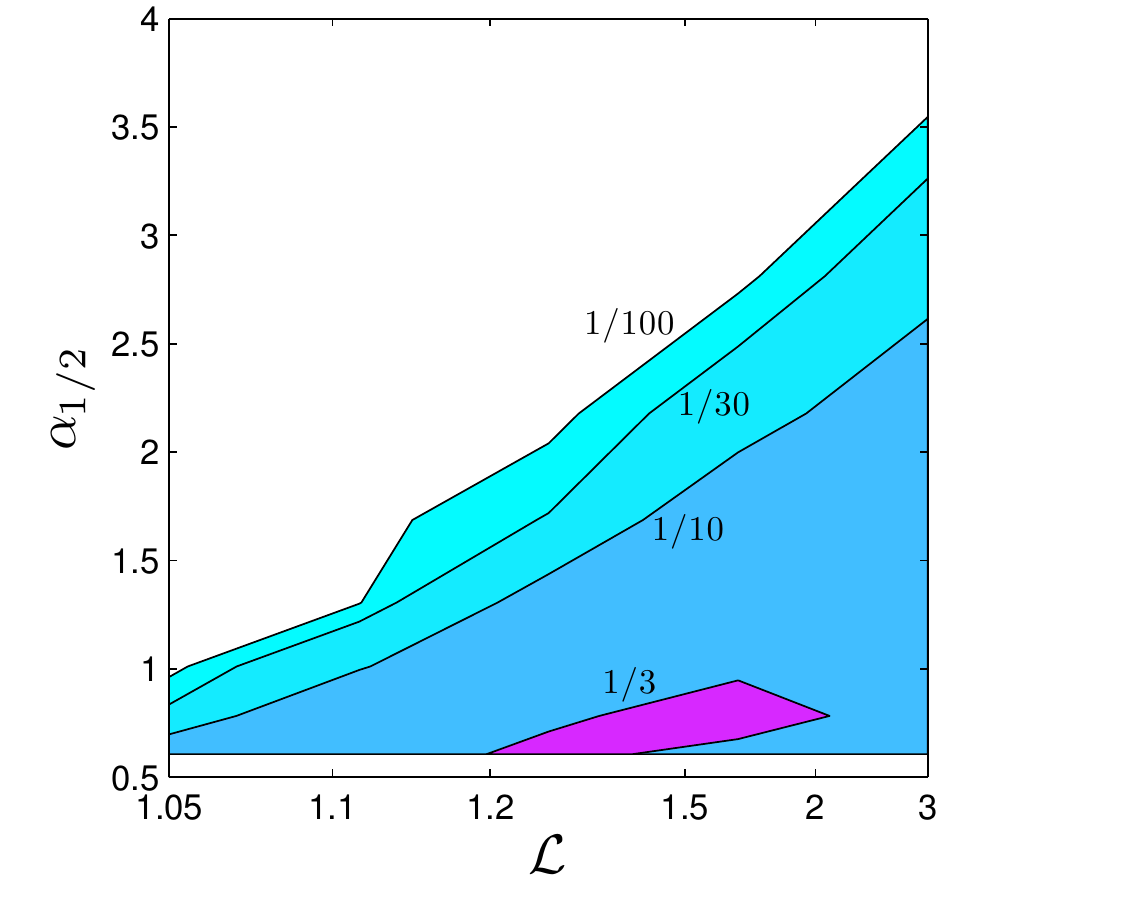}
\caption{ Probability of getting a neutralino LSP as a function of $\alpha_{1/2}$ and $\mathscr{L}$. On both plots, isolines of $\textrm{P}(\tilde{\chi}^0_1\textrm{LSP})$ are indicated
The dependence on $\tan\beta$ and $F^Z/M^*$ is marginal, they are fixed to $\tan\beta=5$, $F^Z/M^*=200~\textrm{GeV}$. 
\textit{Left}: All sign combinations are taken into account with same weight. \textit{Right}: A favorable sign combination.
\label{fig:proba_neut}}
\end{figure}

We therefore compute numerically $\textrm{P}(\tilde{\chi}^0_1\textrm{LSP})$, for all physical sign combinations of the $\lambda$'s appearing in the soft masses.
This probability depends of course on the weights given to the different sign combinations (i.e. the prior). In Figure \ref{fig:proba_neut}, we show   $\textrm{P}(\tilde{\chi}^0_1\textrm{LSP})$ in the $(\alpha_{1/2},\mathscr{L})$ plane. Taking into account all sign combinations, even the one giving tachyons, 
$\textrm{P}(\tilde{\chi}^0_1\textrm{LSP})$ is at most of few percent. If one consider, however, a favorable sign combination, it can   reach $30\%$. 
%It depends mainly on $\mathscr{L}$ and $\alpha_{1/2}$.

\subsection{Flavour constraints}

\begin{figure}
\centering
\includegraphics[width=3.8cm,trim = 0cm 0cm 0cm 0cm, clip]{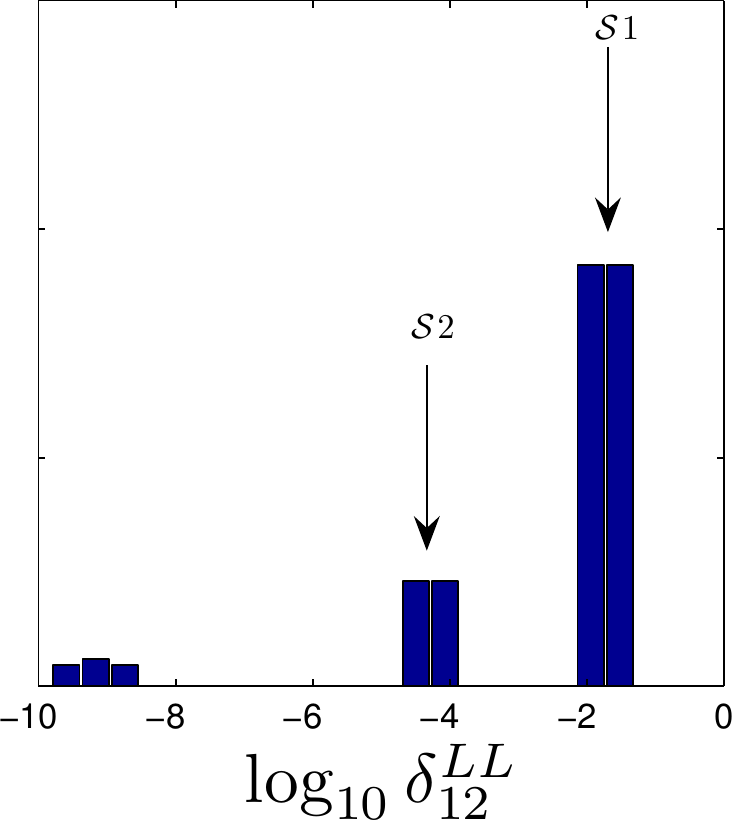}
~~~~~~~~~~~~
\includegraphics[width=3.8cm,trim = 0cm 0cm 0cm 0cm, clip]{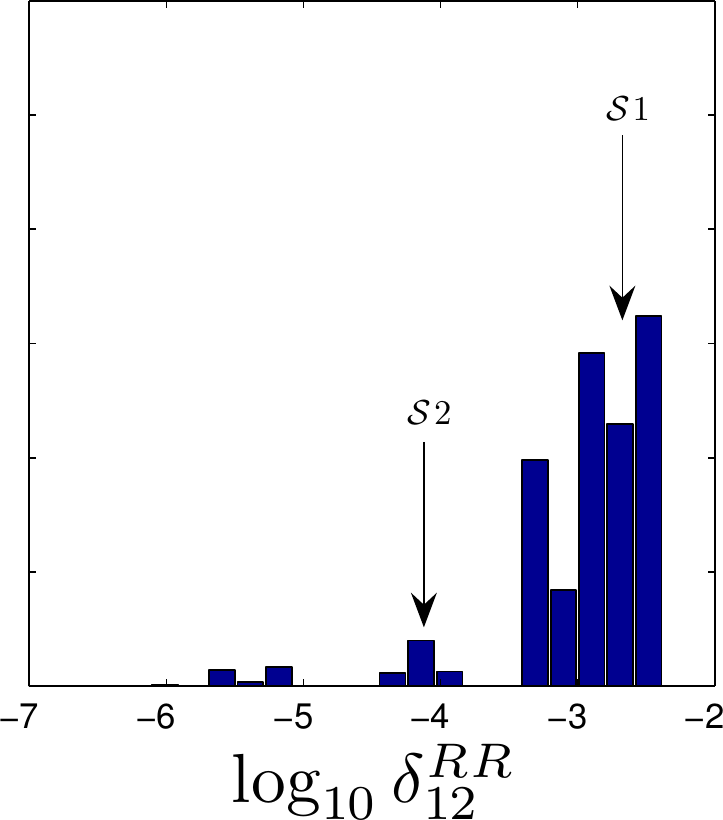}
\caption{ Distribution of mass insertions $\delta^{LL}_{12}$ and $\delta^{RR}_{12}$ for all sign combinations of $\lambda$'s. The model parameters are fixed to $F^Z/M^*=500~\textrm{GeV}$, $\tan\beta=10$, $\alpha_{1/2}=2$, $\mathscr{L}=1$.  The values for the two sign combinations $\mathcal{S}1$ and $\mathcal{S}2$ are indicated by arrows.
\label{fig:sign_dist}}
\end{figure}

Let us finally discuss flavour constraints. This time, not only the mass eigenstates, but also mixings are important. Our strategy is to scan
over all physical sign combinations, keeping $\mathscr{L}=1$, then select representative sign combinations and let vary $\mathscr{L}$.
We focus on lepton flavour violation (LFV). As an example,  we show in Figure \ref{fig:sign_dist} distributions of mass insertions $\delta_{XY}=\frac{\mathcal{M}_{XY}}{\sqrt{\mathcal{M}_{XX}\mathcal{M}_{YY}}}$, for given values of parameters.
Different clusters appear, with more or less suppressed values of $\delta$'s. The origin of these clusters relies on ``accidental" supressions.
 One then has to study how these clusters evolve when  $\mathscr{L}>1$, to check how the accidental supressions survives.
 
% We finally show two representative slices of the parameter space. 
 
\begin{figure}
\centering
\includegraphics[width=6.3cm,trim = 0cm 0cm 11cm 8.3cm, clip]{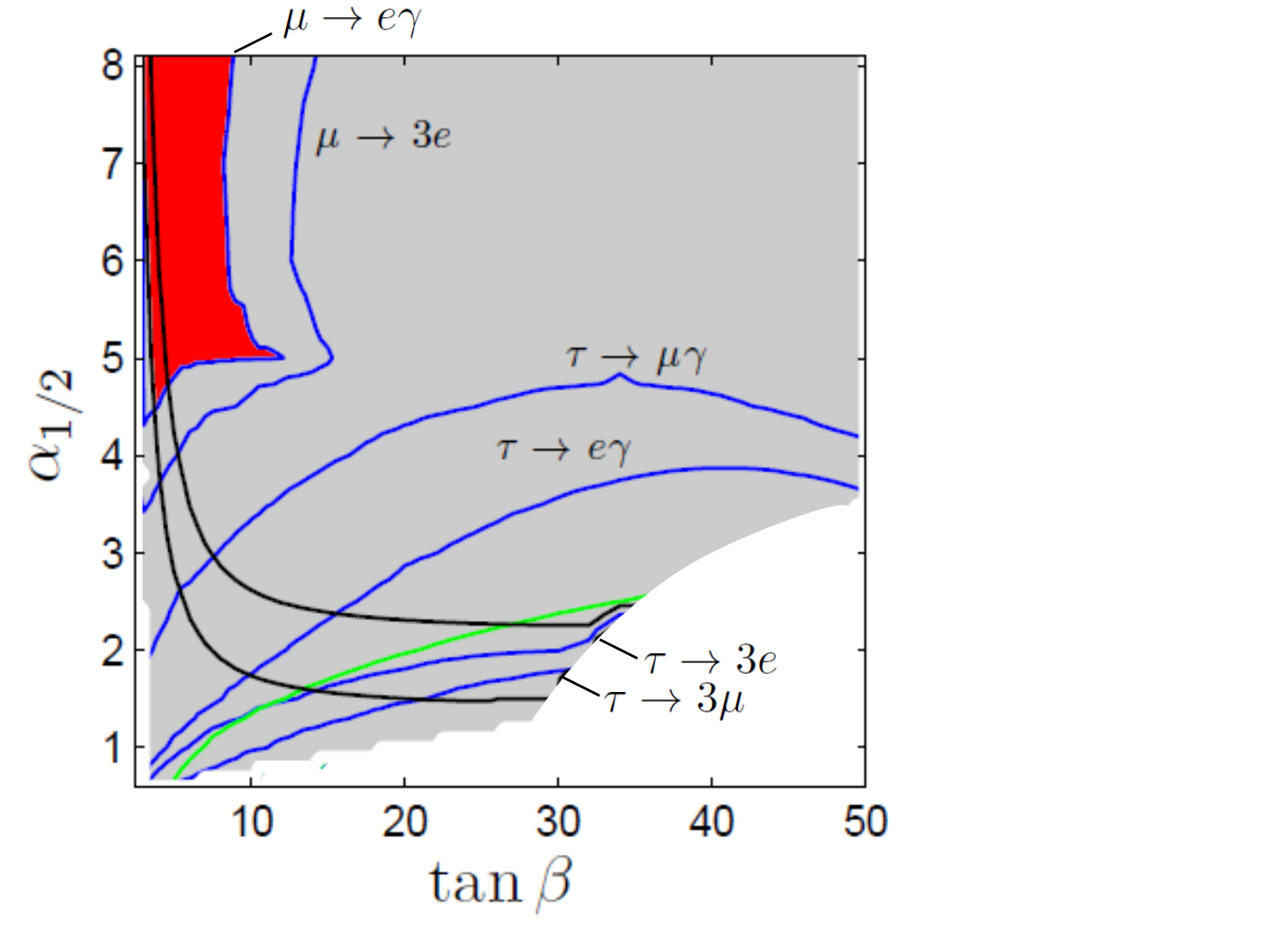}
\includegraphics[width=6.3cm,trim = 0cm 0cm 11cm 8.3cm, clip]{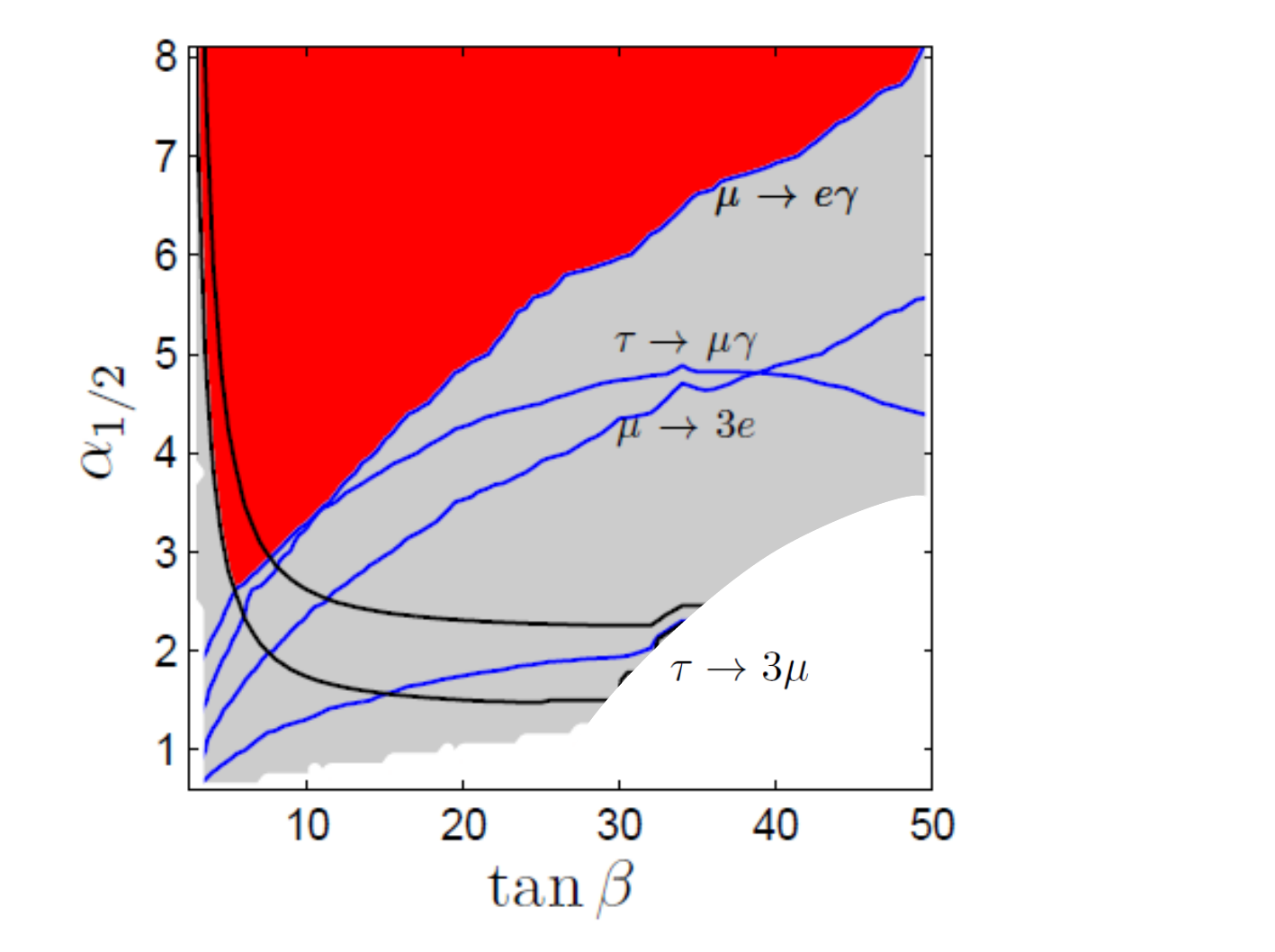}
\caption{ Constraints in the $(\alpha_{1/2},\tan\beta)$ plane, with $\mathcal{L}=1$. The red regions pass all constraints. Blue lines correspond to LFV constraints. The dark lines
correspond to $m_h=111~\textrm{GeV}$ and $m_h=114~\textrm{GeV}$. The green line is a conservative bound on $\Delta a^{BSM}_\mu=450\times 10^{-11}$. \textit{Left}: Sign combination $\mathcal{S}1$.
\textit{Right}: Sign combination $\mathcal{S}2$.
\label{fig:slices}}
\end{figure}

Here, we simply show  in Figure \ref{fig:slices} two slices of the parameter space for $F^Z/M^*\sim 200~\textrm{GeV}$, corresponding to a conservative sign combination $\mathcal{S}1$ and a more favorable sign combination $\mathcal{S}2$.
The most stringent constraints are $\textrm{BR}(\mu\rightarrow e \gamma)<1.2\times10^{-11}$ and the Higgs mass bound. The red regions pass all constraints.
The flavor constraints  weaken if one increases the overall scale $F^Z/M$, as this is the decoupling limit.
 Moreover, depending on $\mathscr{L}$, different mass orderings can appear. 
 In particular,  at this scale, getting a neutralino LSP is highly unlikely. This can be seen by comparing Figure \ref{fig:slices} to Figure \ref{fig:proba_neut}. 
Details will be discussed in \cite{WIP}.

\end{document}